\documentstyle[12pt,qqa4lart]{article}
\input tcilatex

\QQQ{Language}{
American English
}

\begin{document}

\author{Lu-Ming Duan\thanks{%
Electronic address: lmduan@ustc.edu.cn} and Guang-Can Guo\thanks{%
Electronic address: gcguo@sunlx06.nsc.ustc.edu.cn}}
\title{Quantum error avoiding codes verses quantum error correcting codes}
\date{}
\maketitle

\begin{abstract}
\baselineskip 24ptA general theory of quantum error avoiding codes is
established, and new light is shed on the relation between quantum error
avoiding and correcting codes. Quantum error avoiding codes are found to be
a special type of highly degenerate quantum error correcting codes. A
combination of the ideas of correcting and avoiding quantum errors may yield
better codes. We give a practical example.\\

{\bf PACS numbers:} 03.75, 89.70.+c, 03.65.Bz
\end{abstract}

\newpage\ \baselineskip 24ptIn quantum computation or communication, it is
essentially important to maintain coherence of a quantum system [1]. In
reality, however, decoherence due to the interaction of the system with
noisy environment is inevitable [2]. It is discovered that the quantum
redundant coding is the most efficient way to combat decoherence. Until now,
many kinds of quantum codes has been devised [3-24]. According to their
principles, quantum codes can be divided into three classes, i.e., quantum
error correcting codes (QECCs) [3-15], quantum error preventing codes
(QEPCs) [16,17], and quantum error avoiding codes (QEACs) [18-23]. QECCs are
capable of detecting and correcting quantum errors. QEPCs just detect
errors. From the quantum Zeno effect, quadratic noise can be suppressed by
frequent error detections [16,17]. QEACs avoid quantum errors by encoding
input states into coherence-preserving states. These schemes do not need to
detect and correct errors. They are useful with specific noise models
[19,20].

A general theory of QECCs has been established in Ref. [8]. In this paper,
we develop a general theory of QEACs. Necessary and sufficient conditions
for QEACs are obtained. We find an interesting connection between QEACs and
QECCs. QEACs can be regarded as a special type of highly degenerate QECCs,
and furthermore, if degeneracy of a QECC attains the maximum, the code
necessarily becomes a QEAC. The existing QECCs of practical importance all
belong to the class of non-degenerate QECCs [3-15]. These codes are devised
to correct error after occurrence of the error. They do not avoid errors. In
contrast, the QEACs avoid errors, but have no ability to correct errors. A
combination of the ideas of correcting and avoiding errors may yield better
codes. We give examples by devising a class of quantum codes in a practical
circumstance. These codes, which are found to be more efficient than the
QEACs and the existing non-degenerate QECCs, provide interesting examples of
degenerate QECCs.

We start by developing a general formalism for QEACs. A quantum information
system generally consists of many qubits. The system inevitably interacts
with noisy environment. The total Hamiltonian describing the interacting
system is denoted by $H_{tot}$, which may include free evolution of the
qubits, qubit-qubit interactions, and qubit-environment coupling. $H_{tot}$
can be divided into two parts, i.e., 
\begin{equation}
\label{1}H_{tot}=H_c+H_{uc}. 
\end{equation}
The first part $H_c$ refers to the Hamiltonian that is controllable, such as
the free evolution of the qubits and quantum logic operations. The second
part $H_{uc}$ represents the uncontrollable Hamiltonian, such as some noisy
interactions between the qubits and the coupling of the qubits to noisy
environment. In reality, the Hamiltonian $H_c$ makes the system evolve in a
controllable way and results in quantum computation. The Hamiltonian $H_{uc}$
results in noise and decoherence, which should be eliminated if we want to
bring quantum computation into practice.

We discuss the noisy interaction $H_{uc}$ in the interaction picture. The
interaction Hamiltonian has the form (setting $\hbar =1$) 
\begin{equation}
\label{2}H_I\left( t\right) =e^{-iH_ct}H_{uc}e^{iH_ct}. 
\end{equation}
Under this Hamiltonian, after a certain time the reduced density operator of
the system evolves from $\rho _i$ to $\rho _f=\stackrel{\symbol{94}}{S}%
\left( \rho _i\right) $, where $\stackrel{\symbol{94}}{S}$ is the
superoperator associated with the noisy interaction. In the case where the
environment is not initially entangled with the system, $\rho _f$ can be
written in the form [24] 
\begin{equation}
\label{3}\rho _f=\stackunder{a}{\sum }A_a\rho _iA_a^{+}, 
\end{equation}
where the linear operators $A_a$ satisfy the condition 
\begin{equation}
\label{4}\stackunder{a}{\sum }A_a^{+}A_a=I, 
\end{equation}
and $I$ is the unit operator. All the $A_a$ are called interaction
operators. For a given evolution $\stackrel{\symbol{94}}{S}$, the choice of
the operator family $\left\{ A_a\right\} $ is not unique. We choose the
smallest family by requiring that the operators $A_a$ are linearly
independent in the whole Hilbert space of the qubits. Under this
requirement, the number of elements in the family $\left\{ A_a\right\} $ is
uniquely defined, but concrete forms of the $A_a$ are not determined yet.
Look at the following transformation: 
\begin{equation}
\label{5}B_b=\stackunder{a}{\sum }x_{ba}A_a, 
\end{equation}
where the coefficients $x_{ba}$ satisfy the condition $\stackunder{b}{\sum }%
x_{ba}^{*}x_{ba^{^{\prime }}}=\delta _{aa^{^{\prime }}}$, i.e., the matrix $%
X=\left[ x_{ba}\right] $ is unitary. Under the transformation (5), it is
evident that the operator family $\left\{ B_b\right\} $ also satisfies $\rho
_f=\stackunder{b}{\sum }B_b\rho _iB_b^{+}$ and $\stackunder{b}{\sum }%
B_b^{+}B_b=I$. Hence, $\left\{ B_b\right\} $ is also a realization of the
evolution $\stackrel{\symbol{94}}{S}$. In the operator-sum representation
(3), all the operator families linked by the unitary transformation (5) are
equivalent. Because of this equivalence, for a given evolution $\stackrel{%
\symbol{94}}{S}$ we can choose the simplest operator family. In the
following, without loss of generality, we choose that $A_0=\gamma _0I$ with $%
0<\gamma _0<1$. This choice is possible if $\rho _i$ and $\rho _f$ are not
orthogonal to each other. The interaction operators which are not
proportional to $I$ are called error operators.

To give an accurate definition of QEACs, it is convenient to use the notion
of fidelities. For a pure input state $\rho _i=\left| \Psi _i\right\rangle
\left\langle \Psi _i\right| $, which are subjected to noise described by the
operator family $\left\{ A_a\right\} $, the input-output state fidelity is
defined by 
\begin{equation}
\label{6}F\left( \left| \Psi _i\right\rangle ,\left\{ A_a\right\} \right)
=\left\langle \Psi _i\right| \rho _f\left| \Psi _i\right\rangle =\stackunder{%
a}{\sum }\left| \left\langle \Psi _i\right| A_a\left| \Psi _i\right\rangle
\right| ^2. 
\end{equation}
A code $C$ is defined as a subspace of the whole Hilbert space $H$ of the
qubits. The code fidelity is measured by 
\begin{equation}
\label{7}F\left( C,\left\{ A_a\right\} \right) =\stackunder{\left| \Psi
\right\rangle \in C}{\min }F\left( \left| \Psi \right\rangle ,\left\{
A_a\right\} \right) . 
\end{equation}
The code $C$ is defined to be a QEAC if $C$ is a maximal linear subspace of $%
H$ which has the property that the code fidelity 
\begin{equation}
\label{8}F\left( C,\left\{ A_a\right\} \right) =\stackunder{\left| \Psi
\right\rangle \in C}{\min }\stackunder{a}{\sum }\left| \left\langle \Psi
\right| A_a\left| \Psi \right\rangle \right| ^2=1. 
\end{equation}
Suppose that $M$ and $N$ are the dimensions of the code $C$ and of the whole
Hilbert space $H$, respectively. The efficiency of the code is given by $%
\eta =\frac{\log _2M}{\log _2N}.$

The QEAC is characterized by the following theorem:

{\it Theorem 1.} The code $C$\ can be extended to a QEAC iff (if and only
if) $C$\ is a co-eigenspace of all the interaction operators $A_a$.

This condition is more general than the ones given in Refs. [21] and [23],
which are sufficient but not necessary. Theorem1 gives a necessary and
sufficient condition for QEACs.

{\it Proof.} Assume that $C$ can be extended to a QEAC. From the definition
(8), for an arbitrary state $\left| \Psi \right\rangle \in C$, we have 
\begin{equation}
\label{9}\stackunder{a}{\sum }\left| \left\langle \Psi \right| A_a\left|
\Psi \right\rangle \right| ^2=1. 
\end{equation}
The state $A_a\left| \Psi \right\rangle $ can always be decomposed as 
\begin{equation}
\label{10}A_a\left| \Psi \right\rangle =\gamma _a\left| \Psi \right\rangle
+\gamma _a^{\perp }\left| \Psi ^{\perp }\right\rangle , 
\end{equation}
where $\gamma _a,\gamma _a^{\perp }$ are coefficients, and $\left| \Psi
^{\perp }\right\rangle $ denotes a normalized state orthogonal to the state $%
\left| \Psi \right\rangle $. Equations (9) and (10) yield 
\begin{equation}
\label{11}\stackunder{a}{\sum }\left| \gamma _a\right| ^2=1. 
\end{equation}
On the other hand, from Eqs. (4) and (10), it follows that 
\begin{equation}
\label{12}\stackunder{a}{\sum }\left( \left| \gamma _a\right| ^2+\left|
\gamma _a^{\perp }\right| ^2\right) =1. 
\end{equation}
Hence, we have $\gamma _a^{\perp }=0$, i.e., $\left| \Psi \right\rangle $ is
a co-eigenstate of all the interaction operators $A_a$, with the eigenvalues 
$\gamma _a$, respectively. The eigenvalues $\gamma _a$ should be independent
of the state $\left| \Psi \right\rangle $. If $\left| \Psi _1\right\rangle
\in C$ and $\left| \Psi _2\right\rangle \in C$, from the linearity of $C$, $%
c_1\left| \Psi _1\right\rangle +c_2\left| \Psi _2\right\rangle $ also
belongs to $C$. So, $\left| \Psi _1\right\rangle $, $\left| \Psi
_2\right\rangle $ and $c_1\left| \Psi _1\right\rangle +c_2\left| \Psi
_2\right\rangle $ are co-eigenstates of the operators $A_a$. This is
possible iff they have the same eigenvalues; thus $C$ is a co-eigenspace of
all the interaction operators.

The converse of the theorem is straightforward. If $C$ is a co-eigenspace of
all the interaction operators, with the eigenvalues denoted by $\gamma _a$,
respectively, for an arbitrary state $\left| \Psi \right\rangle \in C$,
obviously we have 
\begin{equation}
\label{13}\stackunder{a}{\sum }\left| \left\langle \Psi \right| A_a\left|
\Psi \right\rangle \right| ^2=\stackunder{a}{\sum }\left| \gamma _a\right|
^2=\stackunder{a}{\sum }\left\langle \Psi \right| A_a^{+}A_a\left| \Psi
\right\rangle =1. 
\end{equation}
Hence $F\left( C,\left\{ A_a\right\} \right) =1$, and $C$ can be extended to
a QEAC. This completes the proof of the theorem.

It is interesting to compare QEACs with QECCs. Unlike QEACs, QECCs are
influenced by the error operators. But the influence can be eliminated and
the encoded state can be perfectly recovered by applying an appropriate
recovery operator. From the definition, we see that QEACs can be regarded as
a special type of QECCs, which do not need any recovery operations. Hence, a
QEAC should also satisfy the condition for QECCs. The necessary and
sufficient condition for QECCs has been given in [8]. Assume that $%
C^{^{\prime }}$ is a code, and $\left\{ A_a\right\} $ denotes the family of
interaction operators. The code $C^{^{\prime }}$ can be extended to a QECC
iff for all basisvectors $\left| i_L\right\rangle ,\left| j_L\right\rangle $
of $C^{^{\prime }}$ and operators $A_a,A_b$ in $\left\{ A_a\right\} $%
\begin{equation}
\label{14}\left\langle i_L\right| A_a^{\dagger }A_b\left| j_L\right\rangle
=\gamma _{ab}\delta _{ij}, 
\end{equation}
where the coefficients $\gamma _{ab}$ should be independent of the
basisvectors. The coefficient matrix $\Gamma =\left[ \gamma _{ab}\right] $
is obviously Hermitian, but its form is not uniquely defined, since the
choice of the interaction operators is not unique. All the operator families
linked by the unitary transformation (5) are equivalent. Under these
transformations, the Hermitian coefficient matrix $\Gamma $ can always be
cast into a diagonal matrix, with the eigenvalues being positive real
numbers. If all the eigenvalues of $\Gamma $ do not equal zero, the code is
called a non-degenerate QECC. In contrast, if some eigenvalues equal zero,
or equivalently, if some lines of the coefficient matrix $\Gamma $ are
linearly dependent, the code is degenerate. All the discovered QECCs devised
in practical circumstances belong to the class of non-degenerate codes. A
formal example of degenerate QECCs was given in [8].

Obviously, QEACs should also satisfy the condition (14); but not all the
codes satisfying Eq. (14) are QEACs. What kind of restrictions need be added
for QEACs? The additional restriction is shown by the following theorem,
which provides another form of the necessary and sufficient condition for
QEACs.

{\it Theorem 2.} The code $C$ can be extended to a QEAC iff for all
basisvectors $\left| i_L\right\rangle ,\left| j_L\right\rangle $ of $C$ and
interaction operators $A_a,A_b$ in the family $\left\{ A_a\right\} $%
\begin{equation}
\label{15}\left\langle i_L\right| A_a^{\dagger }A_b\left| j_L\right\rangle
=\gamma _a^{*}\gamma _b\delta _{ij}, 
\end{equation}

{\it Proof.} Assume that $C$ can be extended to a QEAC. From theorem 1, for
an arbitrary $\left| i_L\right\rangle \in C$ and $A_a\in \left\{ A_a\right\} 
$, we have 
\begin{equation}
\label{16}A_a\left| i_L\right\rangle =\gamma _a\left| i_L\right\rangle , 
\end{equation}
thus Eq. (15) holds. Conversely, if Eq. (15) holds, Eq. (4) yields 
\begin{equation}
\label{17}\stackunder{a}{\sum }\left| \gamma _a\right| ^2=1. 
\end{equation}
Suppose that $\left| \Psi \right\rangle $ is an arbitrary state in the
subspace $C$. From Eq. (15), it follows that 
\begin{equation}
\label{18}\left\langle \Psi \right| A_a^{\dagger }A_b\left| \Psi
\right\rangle =\gamma _a^{*}\gamma _b. 
\end{equation}
The choice of the operator family $\left\{ A_a\right\} $ is not unique, and
we can always choose that $A_0=\gamma _0I$. Let $A_b$ in Eq. (17) equal $A_0$%
, then we have 
\begin{equation}
\label{18}\left\langle \Psi \right| A_a\left| \Psi \right\rangle =\gamma _a. 
\end{equation}
The code fidelity 
\begin{equation}
\label{8}F\left( C,\left\{ A_a\right\} \right) =\stackunder{\left| \Psi
\right\rangle \in C}{\min }\stackunder{a}{\sum }\left| \left\langle \Psi
\right| A_a\left| \Psi \right\rangle \right| ^2=\stackunder{a}{\sum }\left|
\gamma _a\right| ^2=1, 
\end{equation}
thus $C$ can be extended to a QEAC. This completes the proof.

Theorem 2 has an interesting corollary. We know that QEACs can be regarded
as special QECCs. It is natural to ask in what circumstances QECCs reduce to
QEACs. This question is answered by the following corollary of theorem 2.

{\it Corollary.} The QECC $C$ reduces to a QEAC iff the coefficients $\gamma
_{ab}$ in Eq. (14) can be decomposed as $\gamma _a^{*}\gamma _b$.

If $\gamma _{ab}$ is decomposed as $\gamma _a^{*}\gamma _b$, the rank of the
coefficient matrix $\Gamma =\left[ \gamma _{ab}\right] $ is not larger than
one, and at most one eigenvalue of the matrix $\Gamma $ does not equal zero.
Hence, in this circumstance the code is highly degenerate. We therefore have
the following conclusion: If degeneracy of a QECC attains the minimum, the
code is a non-degenerate QECC; Conversely, if the degeneracy attains the
maximum, the code becomes a QEAC.

QEACs and non-degenerate QECCs are two extremes. Are there intermediate
circumstances? In the following, we consider a practical decoherence model.
For this model, the optimal code is neither a QEAC nor a non-degenerate
QECC, but a combination of them. The code correct and avoid errors at the
same time, and it provides an interesting example for degenerate QECCs of
practical importance. (To our knowledge, this is the first practical
example).

All the discovered QEACs assume the collective decoherence model [19,20].
Suppose we have $L$ qubits. In the collective decoherence model, the error
operators are described by $A^{+}=\gamma ^{+}\stackunder{l=1}{\stackrel{L}{%
\sum }}\sigma _l^{+}$, $A^{-}=\gamma ^{-}\stackunder{l=1}{\stackrel{L}{\sum }%
}\sigma _l^{-}$, and $A^z=\gamma ^z\stackunder{l=1}{\stackrel{L}{\sum }}%
\sigma _l^z$, where $\overrightarrow{\sigma }_l$ are Pauli's operators. The
three operators $A^{+}$, $A^{-}$, and $A^z$, together with $A_0=\gamma _0I$
make a complete family of the interaction operators. To avoid all the
collective errors, four is the least number of qubits to encode one qubit of
information [20]. In the circumstance of collective decoherence, QEACs are
more efficient than the non-degenerate QECCs. The latter needs at least five
qubits to encode a bit of quantum information [6].

Collective decoherence results form the assumption that the distance between
the qubits is very small so that it is less than the effective wave length
of the noise field [18,19]. It is most possible for the closely spaced
adjacent qubits to satisfy this assumption. Hence, here we assume that every
two adjacent qubits (called a qubit-pair) are decohered collectively; but
the qubits in different qubit-pairs are allowed to decohere in an arbitrary
manner, possibly independently, possibly cooperatively. Suppose that we have 
$2L$ qubits, denoted by $1,1^{^{\prime }},2,2^{^{\prime }},\cdots $, and $%
L,L^{^{\prime }}$, respectively. The $l$ and $l^{^{\prime }}$ $\left(
l=1,2,\cdots ,L\right) $ qubits are decohered collectively. In our
decoherence model, the error operators are described by 
\begin{equation}
\label{20}A_l^\alpha =\gamma _l^\alpha \left( \sigma _l^\alpha +\sigma
_{l^{^{\prime }}}^\alpha \right) , 
\end{equation}
with $l=1,2,\cdots ,L$ and $\alpha =\pm ,$ or $z$. For this decoherence
model, the non-degenerate QECCs need at least five qubits to encode one
qubit of information; and it is impossible to devise any QEACs, for the
co-eigenspace of the error operators $A_l^\alpha $ is of only one dimension
with the sole basisvector $\left| \Psi _i\right\rangle =\otimes _l\left[
\frac 1{\sqrt{2}}\left( \left| 01\right\rangle _{ll^{^{\prime }}}-\left|
10\right\rangle _{ll^{^{\prime }}}\right) \right] $, where $\left|
0\right\rangle _l$ and $\left| 1\right\rangle _l$ are two eigenstates of the
operator $\sigma _l^z$. However, a combination of the ideas of correcting
and avoiding errors can yield better codes. In fact, four qubits are enough
to encode one qubit of information. Suppose that the first qubit $1$ is in
an arbitrary input state $\left| \Psi _i\right\rangle _1=c_0\left|
0\right\rangle _1+c_1\left| 1\right\rangle _1$. The ancillary qubits $%
1^{^{\prime }}$, $2$, and $2^{^{\prime }}$ are prepared in the states $%
\left| 1\right\rangle _{1^{^{\prime }}}$, $\left| 0\right\rangle _2$, and $%
\left| 1\right\rangle _{2^{^{\prime }}}$, respectively. The encoding is
given by the following operation : 
\begin{equation}
\label{21}\left| \Psi _i\right\rangle _1\otimes \left| 101\right\rangle
_{1^{^{\prime }}22^{^{\prime }}}\stackrel{C_{11^{^{\prime
}}}C_{12}C_{12^{^{\prime }}}}{\longrightarrow }\left| \Psi \text{enc}%
\right\rangle =c_0\left| 0101\right\rangle _{11^{^{\prime }}22^{^{\prime
}}}+c_1\left| 1010\right\rangle _{11^{^{\prime }}22^{^{\prime }}}, 
\end{equation}
where all the $C_{ij}$ represent the controlled-NOT (CNOT) operation, with
the first subscript of $C_{ij}$ referring to the control bit and the second
to the target. After this encoding, obviously we have $A_1^z\left| \Psi 
\text{enc}\right\rangle =A_2^z\left| \Psi \text{enc}\right\rangle =0$, where 
$A_l^z$ $\left( l=1,2\right) $ are defined by Eq. (20). Hence, the errors $%
A_1^z$ and $A_2^z$ are avoided. The remaining errors $A_1^{\pm }$ and $%
A_2^{\pm }$ can be easily detected and corrected. We make a quantum
non-demolition measurement of the operators $\sigma _l^z+\sigma
_{l^{^{\prime }}}^z$. If the measurement outcome is $\pm 2$ for an $l$, the
error $A_l^{\pm }$ takes place; and it is readily corrected by performing
some quantum CNOT\ operations. For this decoherence model, the four-bit code
(21) can be easily proven to be optimal by showing that three bits are not
enough to encode one qubit of information.

The above code can be extended straightforwardly to multi-qubit
circumstances. The general input state of $L$ qubits is expressed as 
\begin{equation}
\label{22}\left| \Psi _L\right\rangle =\stackunder{\left\{ i_l\right\} }{%
\sum }c_{\left\{ i_l\right\} }\left| \left\{ i_l\right\} \right\rangle ,
\end{equation}
where $\left\{ i_l\right\} $ denotes $i_1,i_2,\cdots ,i_L$, and $\left|
\left\{ i_l\right\} \right\rangle $ represents $\left| i_1\right\rangle
\otimes \left| i_2\right\rangle \otimes \cdots \otimes \left|
i_L\right\rangle $ with $i_l=0$ or $1$. We use $2L+2$ qubits to encode $L$
qubits of information. The state (22) is encoded into the following state of 
$L+1$ qubit-pairs 
\begin{equation}
\label{23}\left| \Psi _{2L+2}\right\rangle _{\text{enc}}=\stackunder{\left\{
i_l\right\} }{\sum }c_{\left\{ i_l\right\} }\left| \left\{ i_l,\overline{i}%
_{l^{^{\prime }}}\right\} \right\rangle \otimes \left| i_{L+1},\overline{i}%
_{L^{^{\prime }}+1}\right\rangle ,
\end{equation}
where $\left| \left\{ i_l,\overline{i}_{l^{^{\prime }}}\right\}
\right\rangle $ denotes $\left| i_1\right\rangle \otimes \left| \overline{i}%
_{1^{^{\prime }}}\right\rangle \otimes \left| i_2\right\rangle \otimes
\left| \overline{i}_{2^{^{\prime }}}\right\rangle \otimes \cdots \otimes
\left| i_L\right\rangle \otimes \left| \overline{i}_{L^{^{\prime
}}}\right\rangle $ with $\overline{i}_{l^{^{\prime }}}=1-i_l$. The state $%
\left| i_{L+1},\overline{i}_{L^{^{\prime }}+1}\right\rangle $ of the $(L+1)$%
th qubit-pair should satisfy $\stackrel{L+1}{\stackunder{l=1}{\sum }}i_l=0$
mod $2$. Through the encoding (23), all the errors $A_l^z$ are avoided. The
remaining errors $A_l^{\pm }$ can be detected and corrected by the procedure
very similar to that in the single qubit circumstance. For a large $L$, the
efficiency of the code is approximately $\frac 12$. In the case of $L\geq 2$%
, the code (23) is not necessarily optimal; but it has the advantage of
being very simple, and easy to encode, decode, and to detect the error
syndrome.

All the discovered QEACs assume the collective decoherence model. Before
ending the paper, we emphasize that QEACs may also find their application in
other decoherence models. For example, it is possible to avoid some
correlated errors by QEACs. As a simple example, we assume that there are
only two qubits, subject to the following correlated errors. The error
operators are given by $A_1=\sigma _1^{+}\sigma _2^{-}$, and $A_2=\sigma
_2^{+}\sigma _1^{-}$. These errors can be avoided by the encoding 
\begin{equation}
\label{24}
\begin{array}{c}
\left| 0\right\rangle \rightarrow \left| 00\right\rangle , \\  
\\ 
\left| 1\right\rangle \rightarrow \left| 11\right\rangle .
\end{array}
\end{equation}
One qubit of information is encoded. This simple example suggests that QEACs
may have wide use. It is an interesting question to find further
applications of QEACs in other practical decoherence models.\\

{\bf Acknowledgment}

This project was supported by the National Nature Science Foundation of
China.

\newpage\

\end{document}